\documentclass{svproc}
\usepackage{url}

\usepackage[pdftex]{graphicx}
\usepackage{cite}

\begin{document}
\mainmatter              
\title{Visual Response to Emotional State of User Interaction}
\titlerunning{Mood spRing}  
%
\author{Nina Marhamati\inst{1} \and Sena Clara Creston\inst{2}
}
\authorrunning{Nina Marhamati et al.} 
%
\tocauthor{Nina Marhamati, Sena Clara Creston}
\institute{Computer Science Department, Sonoma State University,
Rohnert Park, CA 94928,\\
\email{nina.marhamati@sonoma.edu}
\and
Department of Art, Sonoma State University,
Rohnert Park, CA 94928,\\
\email{crestons@sonoma.edu}}

\maketitle              

\begin{abstract}
This work proposes an interactive art installation ``Mood spRing" designed to reflect the mood of the environment through interpretation of language and tone. Mood spRing consists of an AI program that controls an immersive 3D animation of the seasons. If the AI program perceives the language and tone of the users as pleasant, the animation progresses through idealized renditions of seasons. Otherwise, it slips into unpleasant weather and natural disasters of the season. To interpret the language and tone of the user interaction, hybrid state-of-the-art emotion detection methods are applied to the user audio and text inputs. The emotional states detected separately from tone and language are fused by a novel approach that aims at minimizing the possible model disparity across diverse demographic groups.
\end{abstract}

\section{Introduction and Background}
With the growth of cyberbullying and its impact on artificial intelligence (AI) projects, such as Tay \cite{tay}, there is an urge to help public understanding of and interaction with AI. In \cite{turkle2006relational}, the authors use the term ``robot as Roschack" to explain how “relationships with robots express other things about a person’s life”. Mood spRing is created as an interactive art installation to give viewers creative freedom to express themselves in an environment that mirrors their emotional output. This emotional reflexivity is created to enable the viewer to feel how their language and tone make others feel.
	
This connection between the action of the viewer, the reaction of the artwork, and the emotional response this relationship nurtures has been pioneered by other artists. 
The software artist Erik Loyer explores the connection between the viewers’ actions and environmental reactions in his 2013 app Breathing Room \cite{broom} in which the viewer’s movements mimic the wind to trigger environmental and anecdotal responses. 

Incorporating the viewer’s responses into the artwork brings attention to the value of their actions. The artist, Sena Clara Creston, uses interactive artwork to enable viewers to experience the effects of their actions past the scope of their observation. Her previous interactive 3D environment, The Cloud Factory \cite{cf} enables viewers to be the creator and destroyer of their environment, promoting reflection on one’s constructive and destructive power. Her other installations utilize familiar discarded materials and physical positive and negative responses to encourage the viewer to see the consequences of their actions. Creating a responsive environment cites the power of the viewer’s autonomous actions to control a response, with the intention of asking the viewer to consider their environmental impact.

To create such a response, we use AI approaches to detect the emotional state of the user interaction. Human emotions are complicated, however, several psychological models of affect have been used by the AI community for emotion detection, such as Russel’s circumplex and Plutchik’s wheel of emotion \cite{acheampong2020text}. Most existing emotion detection datasets use six basic categories of happiness, sadness, anger, disgust, surprise, and fear to tag emotions  \cite{acheampong2020text}. On the other side, psychologists have pointed out the expression of emotions varies across different cultures and individuals \cite{barrett2019emotional} and only pleasant and unpleasant emotions could be universally recognized \cite{barrett2017emotions} . In this work, we propose to use Russel’s circumplex to map the six emotion categories to pleasant and unpleasant classes. Using these two classes makes our methodology generalizable to different cultures and individuals and better facilitates the mechanism for controlling the visual response.

The AI research community has tackled emotion detection based on the type of data. Text-based approaches could be categorized to keyword-based, rule-based, classical machine learning such as SVM, deep learning, and hybrid approaches \cite{alswaidan2020survey}. There are over a dozen of benchmark datasets for text-based emotion detection  \cite{acheampong2020text}, such as ISEAR \cite{scherer1994evidence}, the Daily Dialog \cite{li2017dailydialog}, and EmoBank \cite{buechel2017emobank}.

Detecting emotions from speech has been performed by extracting prosodic features to represent the auditory qualities of sound such as pitch, and spectral features, which are frequency-based \cite{sailunaz2018emotion}. Mel frequency cepstrum coefficients (MFCC), linear prediction cepstral coefficients (LPCC), and their variants are some of such commonly used features \cite{jain2018cubic}. Classifiers based on support vector machines (SVM), Gaussian mixture model (GMM), and hidden Markov model (HMM) have been applied to those features for emotion detection \cite{zamil2019emotion}. Deep learning approaches have had great success classifying the audio features \cite{venkataramanan2019emotion} or generating their own embeddings and features directly from the audio inputs by training large deep models such as yamnet and VGGish \cite{45857, 45611} and by transfer learning from pre-trained models to classify images of spectrograms \cite{lech2020real}. High dimensional feature space and limited datasets
are some of the challenges of detecting emotions directly from audio inputs. 

\section{Expected Outcome}
Illustrating emotional response to spoken input is expected to encourage viewers to consider the impact of their actions, language, and tone; promoting empathy and understanding across language, cognitive, and technological barriers. As the viewer enters the digital space and receives visual environmental feedback for their perceived input, they may first experiment with the boundaries of the application; Viewers may act on the spectrum from being timid and sensitive, or brazen and aggressive. The more time the viewer spends with the app talking to others without considering the outcome, the more likely the app is going to catch them unaware, unbiasedly reflecting their natural language and tone. Regardless of how the viewer engages, articulated environmental reactions will reinforce the lasting impact of their actions. 

The proposed methodology enables such environmental reactions and investigates improvements in emotion detection when information fusion is used. Utilizing multiple sensors in parallel and the addition of an extra layer for information fusion is expected to reduce gender/age bias and improve the overall results. 

\section{The Significance of Experiencing Interactivity Across Digital Forms}
Initially, Mood spRing will be formatted as an app to be experienced on a computer, or a mobile device, enabling maximum participation in various situations. Environmental allusions to creating a 3D animation of changing weather suggest a benefit of finalizing Mood spRing in an immersive virtual reality format. While this format provides optimal immersion, virtual reality might prohibit viewers without the proper hardware from being able to experience it. Incorporating a virtual reality headset may further socially isolate the viewer, discouraging social communication, and free use of language and tone. As an art exhibition, Mood spRing could be displayed as an immersive video installation with four projected walls. This installation would enable people to congregate in the gallery while communicating freely. A few people talking calmly would be reflected by a calm relaxing environment, while a gallery filled with a cacophony of conversation will be reflected by a tumultuous environment. A populated gallery setting might provoke viewers to test or tease the environment to elicit maximum response, disregarding emotional feedback. Alternatively, providing intermediary indicators of an emotional barometer as a response to spoken language and tone has the potential to provide technical applications as emotional indicators for people with a need for assistance interpreting emotional indicators. Other possible applications are enhancing user experience by reflecting the mood of the room in online learning, augmented reality, and game environments.

\section{Methodology}
The majority of the studies on emotion detection (ED) use either plain text or merely audio recordings to detect emotions \cite{sailunaz2018emotion, acheampong2020text, alswaidan2020survey}. Authors of \cite{popovic2020automatic} refer to multi-view representation learning as an approach to combine the result of ED from speech with the result of text-based ED extracted from automatic speech recognition (ASR) system. However, they do not actually apply that to their implementation pipeline. There are other major efforts in emotion recognition from videos, mostly by combining visual and audio inputs \cite{song2018decision}.

There are limited benchmark datasets available on detecting emotions directly from speech \cite{burkhardt2005database, livingstone2018ryerson} which affects the performance of the models trained on such limited data. Models trained on the aforementioned datasets usually focus on the frequency content of the input voice and extract their features from the frequency domain. On the other hand, there have been many more datasets and studies on ED from text streams, such as tweets \cite{hasan2019automatic}, comments, messages, etc. \cite{acheampong2020text, alswaidan2020survey}. Most ED approaches process visual inputs using facial features, which also has been proven ineffective \cite{barrett2019emotional}. The language content of speech carries the semantics and might be crucial in detecting the emotional state of an interaction however that can suffer from gender bias in predicting the results \cite{sun2019mitigating}. As mentioned earlier, models trained on speech only are not very reliable due to limited existing data. Therefore, none of these can be a reliable source of detecting emotions individually.
\begin{figure}[!t]
    \centering
    \includegraphics[scale=.5]{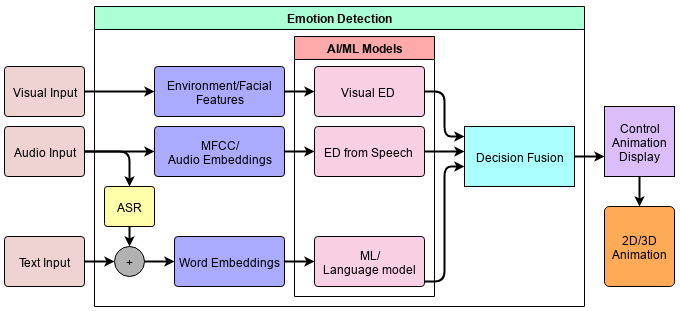}
    \caption{The schematic of emotion detection for generating visual response to human interaction in Mood spRing}
    \label{fig:schematic}
\end{figure}


The methodology proposed here combines the ED results from text with the results from audio and visual inputs. The schematic of the proposed methodology is shown in Fig. \ref{fig:schematic}. The ED system receives different types of inputs from the user, detects the emotional state of the interaction for each input, fuses the predictions from individual models, and outputs the detected emotional state. The output of the ED system is used to generate a control signal to display the animation as a response to the user. We consider either text input from the user or we generate text from the audio input utilizing off-the-shelf automatic speech recognition (ASR). Features are extracted from the inputs by transfer learning from pre-trained models. From here on in the paper, for sake of brevity and without loss of generality, we explain the details of the methodology for audio and text inputs. However, our approach is extensible to visual inputs as well.

For obtaining the audio embeddings, yamnet and VGGish models \cite{45857, 45611} and for the word embedding, the BERT language model \cite{devlin2018bert} are employed here. It is possible to investigate fine-tuning of the same models to detect emotions from each input. In order to provide real-time response to the user, the proposed method applies lightweight models such as support vector machine (SVM), k-nearest neighbors (KNN), and naive Bayes (NB) for ED. Thus, we use other features such as the mel-frequency cepstrum coefficients (MFCC) for audio and term frequencies or TF-IDF values for text as well. 

For decision fusion, classic approaches such as majority voting or ensemble methods for decision-making \cite{polikar2006ensemble} can be used. For improved decision fusion, we propose to train a simple model, such as a small neural network of a fully connected layer, with a fairness measure as the loss function. The fairness measure is designed based on Bernstein bounds \cite{ethayarajh2020your}. Our proposed fusion method weighs the result of each classifier and fuses them such that gender bias is minimized in the final decision. Gender bias is targeted here due to its presence in the RAVNESS dataset. However, this proposed approach is applicable for minimizing bias against any other groups as well. To the best of the authors’ knowledge, there has not been any previous decision fusion approach for minimizing unfairness against underrepresented groups.

The probability of the predicted emotional state of the user interaction is used to control the timing of the animation. Other aspects of the AI model prediction, such as the confidence level can be reflected by controlling the brightness of the animation.

\section{Conclusion and Future Direction}
Mood spRing is the initial collaboration between an AI expert and an artist exploring how programmed reactions to viewer’s actions promote empathetic emotional awareness. Relational programs can be used for future art installations which focus on animating environmental responses to the actions of the viewer, encouraging the viewer to understand the effect their actions have on the greater world. To scale the proposed methodology to such environments with multiple users and multi-modal inputs, we will use benchmark datasets such as MELD \cite{poria2018meld} and apply multi-view representation learning approaches such as canonical correlation analysis \cite{li2018survey}. These programmed relationships can also be used for game development to support relationships between virtually distanced participants.

%
%



\bibliographystyle{spphys.bst}
\bibliography{references}
\end{document}